\documentclass[twocolumn,twoside,slac_two]{revtex4}
\usepackage{graphicx}
\usepackage{fancyhdr}
\def\bi {\begin{itemize}}
\def\ei {\end{itemize}}

\def\psrb{PSR~B1259--63}

\def\deg {$^\circ$}
\def\gr{$\gamma$-ray}
\def\lsi {LSI~+61\deg~303}
\def\ls {LS~5039}

\begin{document}

\title{Study of gamma-ray loud binaries in the Fermi era}

%

\author{M. Chernyakova}
\affiliation{DIAS, 31 Fitzwilliam Place, Dublin 2, Ireland}
\author{A. Neronov}
\affiliation{ISDC, 16 Chemin d'Ecogia, Versoix, Switzerland}

\author{M. Ribordy}
\affiliation{EPFL, 1015, Ecublens, Switzerland}

\begin{abstract}
We discuss {\it Fermi} observations of \gr\ loud binaries. We show that within hadronic model of activity of \lsi, detection of cut-off in the GeV spectrum constrains maximal energy of the  primary protons. In this way, the GeV \gr\ data impose constraint on the expected neutrino signal (spectrum and lightcurve) from the source. We also briefly discuss perspectives of  GeV band detection of \psrb\ during the 2010 periastron passage. 
\end{abstract}

\maketitle

\thispagestyle{fancy}


\section{Introduction}

Gamma-ray-loud binary systems (GRLB) are X-ray binaries  which emit very-high
energy (VHE) \gr s. Four such systems \psrb, \ls, \lsi\ and HESS J0632+057, have 
been firmly detected as persistent or regularly variable TeV \gr\
emitters  \cite{aharon05,aharon06a,albert06, aharon06b,acciari08}. Most of the variable and transient  Galactic sources of GeV \gr s are expected to belong to the GRLB class. 

The source of the high-energy activity of GRLBs is uncertain. It can be either accretion onto  or dissipation of rotation energy of the compact object. It is commonly assumed that the \gr\ emission is produced in result of interaction of the relativistic outflow from the compact object with the non-relativistic wind and radiation field of companion massive star. Neither the nature of the compact object (black hole or neutron star?) nor the geometry (isotropic or anisotropic?) and physical properties (electron-positron or electron-proton composition?) of relativistic wind from this compact object  are known in in most of the GRLBs. The only exception is \psrb\ system in which the compact object is known to be a young rotation powered pulsar which produces relativistic pulsar wind. 

Two GRLBs, \lsi\ and \ls\ were found to be strong sources of 0.1-10~GeV \gr s \cite{abdo09b,abdo09c}. In this contribution we discuss constraints on the physical model of one of these sources, \lsi, imposed by  {\it Fermi} observations. We demonstrate that {\it Fermi} observations significantly constrain the expected neutrino signal from the source, within hadronic model of source activity. For  \psrb, which is not (yet) detected by {\it Fermi}, we discuss the expected GeV emission from the source during the next passage of the young pulsar close to the massive star, due in 2010.

\section{GeV~spectral cut-off in \lsi}

\begin{figure}
\includegraphics[width=\linewidth]{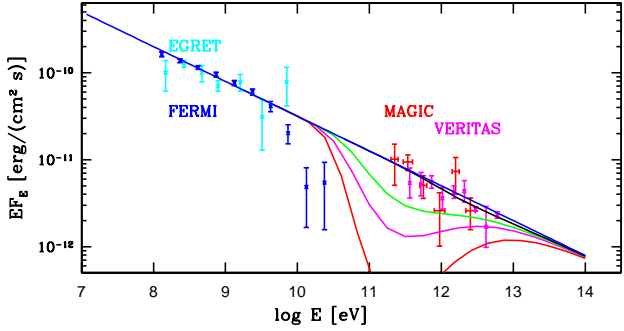}
\caption{Attenuation of \lsi\ spectrum due to the $\gamma\gamma$ pair production at different orbital phases. Blue line is an extrapolation of a GeV spectrum before the cut-off. Black, and red lines shows the effect of $\gamma$-ray absorption at inferior and superior conjunctions, while green and magenta lines correspond to intermediate ($\varphi$= 0.6 and 0.8) phases.}
\label{abs}
\end{figure}

LAT spectrum of \lsi\ in sub-GeV domain is described by a power law $dN_\gamma/dE\sim E^{-\Gamma}$ with a photon index $\Gamma=-2.4$ \cite{abdo09b}. This slope is consistent with low-energy extrapolation of the MAGIC and VERITAS spectra \cite{albert06,acciari08}, see Fig. \ref{abs}.  LAT reveals a suppression or cut-off in the spectrum above $\sim 6$ GeV  \cite{abdo09b}. In principle, suppression of the source flux in 10~GeV -- 1~TeV band is expected, because \gr\ flux in this energy band is strongly affected by the effect of  pair production on the ultra-violet photon field of companion star  \cite{dubus06}.  Fig.\ref{abs}  shows the attenuation of \lsi\ spectrum due to the $\gamma\gamma$ pair production at different orbital phases under the assumption that the TeV flux comes directly from the compact object.  Strongest attenuation of the source flux occurs close to the periastron /  inferior conjunction of the orbit. 

It is clear from Fig. \ref{abs} that  if the observed GeV-TeV emission originates from the vicinity of the compact object, the pair production optical depth along the line of sight is too small to explain the suppression of the source flux  at several GeV.  If \gr\ emission comes from an extended region or interaction of relativistic wind/jet with the wind of companion star, the emission region could be situated closer to the Be star surface leading to stronger attenuation of \gr\ flux.  Although account of spatial geometry of \gr\ emission region could increase the suppression of the source flux above $\sim 10$~GeV, it could hardly affect the flux at the energies below $10$~GeV. Indeed, \gr s with energies $E_\gamma\lesssim 10$~GeV could interact only with soft photons with energy $\epsilon_{UV}\gtrsim 25\left[E_\gamma/10\mbox{ GeV}\right]$~eV, which is much larger than the typical energy $\epsilon_*\simeq 3kT_*\simeq 7\left[T_*/3\times 10^{4}\mbox{ K}\right]$~eV of the thermal stellar photons  with temperature $T_*$. 

Simple "powerlaw modified by absorption" model of the source spectrum is in contradiction with the observed orbital modulation of the signal in the 0.1-1~TeV energy band. {\it Fermi} orbital lightcurve exhibits a pronounced maximum around $\phi\simeq 0.4$. Fig. \ref{abs} shows that the effect of $\gamma\gamma$ pair production at this orbital phase is small, so that a maximum at the phase $\phi\sim 0.3-0.4$ is expected also in the 0.1-1~TeV band lightcurve. To the contrary, MACIC and VERITAS observe the maximum of emission at the orbital phase $0.6<\phi<0.8$, at which the signal should be suppressed  \cite{albert06,acciari08}.

\begin{figure}
\includegraphics[width=\linewidth]{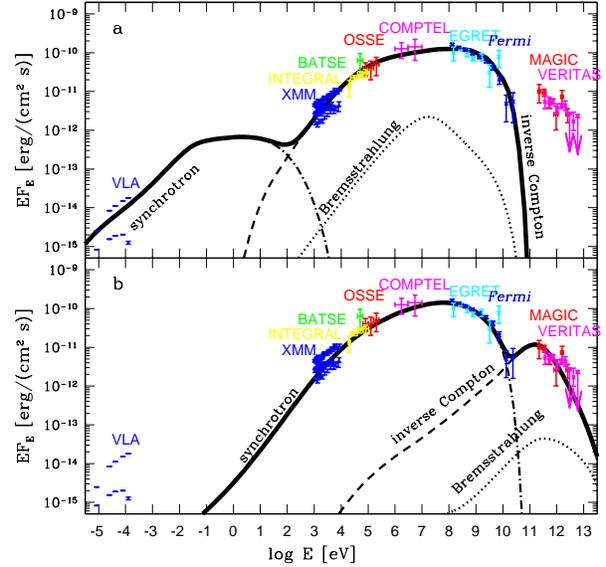}
\caption{Model spectra compared to the data, which are the same as those in \cite{chernyakova06} except for the added data from the VERITAS \cite{acciari08} and {\it Fermi} \cite{abdo09b} telescopes. The dot-dashed, dotted and dashed curves show the spectral components from the synchrotron, bremsstrahlung and IC processes, respectively. Adopted from \cite{zdz09}.}
\label{lsi_synch}
\end{figure}

Since the effect of $\gamma\gamma$ pair production can not explain the cut-off at GeV energies and the difference in the orbital modulation of the source flux at GeV and TeV energies, 
one has to consider a possibility that GeV and TeV \gr s are produced via different mechanisms (e.g. synchrotron and inverse Compton) and/or by different particle populations (e.g. electrons and protons). Fig. \ref{lsi_synch}, adopted from the Ref. \cite{zdz09}, shows two possible models of broad-band spectrum, consistent with {\it Fermi} data.  In the upper panel the GeV band emission is inverse Compton emission from electron distribution with a high-energy cut-off at $E_{cut}\simeq 10$ GeV.   Alternatively, in the model shown in the lower panel of Fig. \ref{lsi_synch}, the GeV band emission is synchrotron emission from electron distribution extending to the energies $E_e\sim 100$~GeV (see \cite{zdz09}).

In the synchrotron model of GeV emission, shown in the lower panel of Fig. \ref{lsi_synch},  the electron injection spectrum has to be very hard \cite{neronov09,zdz09}.  Such hard electron spectrum could be produced if electrons are directly injected from relativistic pulsar wind  \cite{zdz09} or produced in interactions of high-energy protons \cite{neronov09}. 

\section{Orbital modulation of neutrino flux}

\begin{figure}
\includegraphics[width=0.7\linewidth]{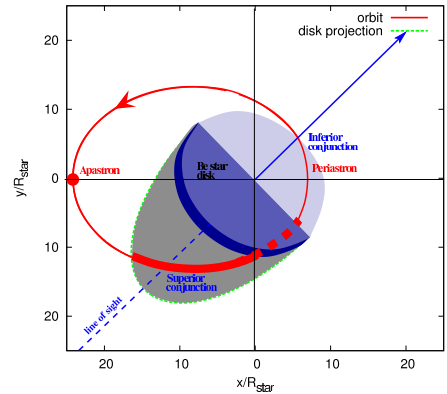}
\caption{Illustrative scheme for the restricted time of neutrino emission. Grey shaded region shows the "shadow" of the Be star disk on the orbital plane, from the viewpoint of observer.}
\label{fig:scheme}
\end{figure}

If high-energy electrons emitting GeV synchrotron radiation are injected via interactions of high-energy protons,   measurement of a cut-off energy of {\it Fermi} spectrum $E_{\gamma,cut}\simeq 6$~GeV \cite{abdo09b} imposes a constraint on the high-energy cut-off in the spectrum of the secondary $e^+e^-$ pairs, $E_{e,cut}\simeq 10^{14}\left[B/10\mbox{ G}\right]^{-1/2}\left[E_{\gamma,cut}/6\mbox{ GeV}\right]^{1/2}$~eV, where $B$ is the magnetic field strength in the emission region. This, in turn, constrains the cut-off in the spectrum of primary protons, 
$E_{p,cut}\simeq \kappa^{-1}E_{e,cut}\simeq 0.5\left[B/10\mbox{ G}\right]^{-1/2}\left[E_{\gamma,cut}/6\mbox{ GeV}\right]^{1/2}\mbox{ PeV}$
where $\kappa\simeq 0.2$ is the typical inelasticity of $pp$ collisions. 

PeV protons do not interact efficiently with the UV photons from the Be star, since the center of mass energy of proton-photon collisions is below the threshold of pion production $E_{p\gamma}\simeq  2\times 10^{16}\left[\epsilon_*/10\mbox{ eV}\right]$~eV. Therefore {\it Fermi} data favor proton-proton hadronic models as compared to proton-photon interaction models, unless the magnetic field in the interaction region is at the level of $B\le 0.1$~G.

The synchrotron interpretation of the source spectrum in the keV-GeV energy range enables to deduce the shape of the electron spectrum from the measured shape of the synchrotron spectrum. This was done in the Ref. \cite{neronov09} in which it was found that the data are consistent with the assumption that $e^+e^-$ pairs are initially injected with energies close to the high-energy cut-off. Hard injection spectrum of $e^+e^-$ pairs could be explained  by the initial hard (much harder than $E^{-2}$) injection spectrum of protons (e.g. injection from the proton-loaded relativistic wind with large bulk Lorentz factor). The spectrum of neutrinos produced in interactions of high-energy protons is expected to be sharply peaked at the highest energies $E_\nu\simeq \kappa E_p\sim 10^{14}\left[B/10\mbox{ G}\right]^{-1/2}\left[E_{\gamma,cut}/6\mbox{ GeV}\right]^{1/2}$~TeV.

Neutrino flux is expected to be modulated on the orbital time scale. Accounting for the orbital periodicity of the neutrino signal significantly increases the IceCube potential for the detection of the source, once the validity of the model prediction is assumed. The resulting model-dependent approach in search for a neutrino signal is diametrically opposed to a random search for steady neutrino point source over the whole sky and complementary to a search for steady neutrino point source selected out of a catalogue~\cite{Abbasi:2009iv}.

The neutrino flux is not expected to be positively correlated with the TeV emission, which is affected by the orbital phase dependent absorption in interactions with UV photons from Be star (see above). It is also not expected to be correlated with the GeV band \gr\ emission, which  might be strongly affected by development of electromagnetic cascade initiated by high-energy proton interactions. Thus, neutrino lightcurve has to be calculated independently, rather than read out from photon lightcurve.

PeV protons are able to penetrate deep into (or even through) the dense equatorial disk of Be star, because their Larmor radius, $R_L\simeq 3\times 10^{12}\left[E_p/10^{15}\mbox{ eV}\right]\left[B/1\mbox{ G}\right]^{-1}$~cm, is comparable to the thickness of the disk. Since proton trajectories are not randomized by the magnetic fields in the disk, the neutrino emission from the system is expected to be anisotropic, with the maximum flux emitted into a cone with the axis aligned with the compact object -- Be star direction \cite{neronov09}. A neutrino signal could be detected if the neutrino emission cone passes through the line of sight, i.e. if the compact object is situated behind the dense equatorial disk of Be star during a part of the orbit, as it is shown in Fig. \ref{fig:scheme}.

\begin{figure}
\includegraphics[width=0.9\linewidth]{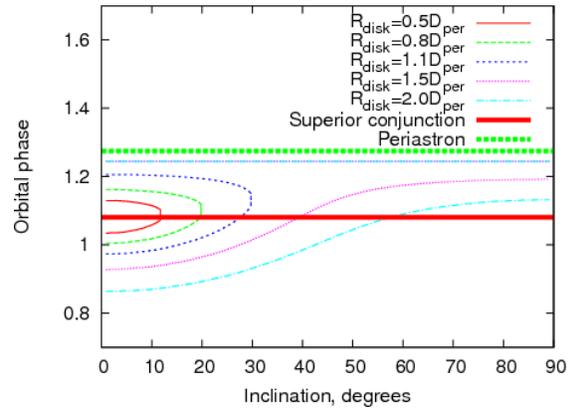}
\caption{The thick lines represent the orbital phase of the superior conjunction and periastron. The thin curves show the range of orbital phase with an expected neutrino emission for different sizes of equatorial disk around  Be star.}\label{fig:phase}
\end{figure}

We consider a disk which is truncated at the distance $R_{\rm disk}$ comparable to the binary separation of the star and compact object at the periastron \cite{zdz09}. The span of the orbital phase period during which the compact object is situated behind the disk is maximized if the disk is oriented perpendicularly to the line of sight, as it is shown in Fig. \ref{fig:scheme}. For such a disk orientation, the result of calculation of the interval of orbital phases, during which the compact object is in the disk shadow (i.e., the interval during which neutrino signal from the source could be detected), is shown in Fig. \ref{fig:phase}. For this calculation, we use the estimates of the orbital parameters of the system from Ref. \cite{aragona09}.

 The interval of phases at which neutrino signal is expected, depends on the (unknown) inclination of the binary orbit.  The width of the phase interval of neutrino emission increases with the increase of the disk size. In Fig. \ref{fig:phase} we show the results of calculation of the phase intervals of neutrino emission for a range of disk sizes, from $R_{\rm disk}$ equal to the half of the binary separation distance at periastron $D_{\rm per}$ to twice the binary separation at periastron. 

It is interesting to note, that the equatorial disk of Be star in the \lsi\ system is observed to periodically vary on  super-orbital time scale of $T_{\rm so}\simeq 4.6$~yr \cite{gregory02}. This means that the orbital phase interval of neutrino emission is also expected to be modulated on the 4.6~yr time scale. The strongest neutrino signal (corresponding to the largest span of the neutrino emission phase interval) is expected when the equatorial disk expands to its maximal size, roughly once in each 4.6~yr cycle. 

It is clear from Fig. \ref{fig:phase} that, independently of the disk size and inclination of the orbit, the phase interval of neutrino emission from the source is $\delta\phi\lesssim 0.3$. If this neutrino emission model within a reduced time window closely matches the actual source neutrino emission, the sensitivity of a neutrino telescope for the detection of the source is increased due to (1) an atmospheric neutrino reduction by a factor $\delta\phi^{-1}\gtrsim 3$ and (2) beyond a free phase variability fit due to the constraint of the phase of the neutrino emission.

\section{Origin of the \psrb\ emission.}

\psrb\ is the only \gr\ loud binary system for which the nature of the compact object and the source of the power are certain. It is clear that (unpulsed) radio-to-TeV emission from the system is generated in collision of relativistic pulsar wind with the dense equatorial wind of Be star.  The line of intersection of the disk plane and the orbital plane is tilted at  about $90$\deg\ with respect to the major axis of the binary orbit, so that the pulsar passes through the disk twice  per orbit. Episodes of pulsar passage through the disk are associated with pronounced flares of the source, visible in radio, X-ray and TeV \gr\ bands \cite{aharon05,tavani97,chernyakova06,uchiyama09,chernyakova09}. 

\begin{figure}
\includegraphics[width=0.9\linewidth]{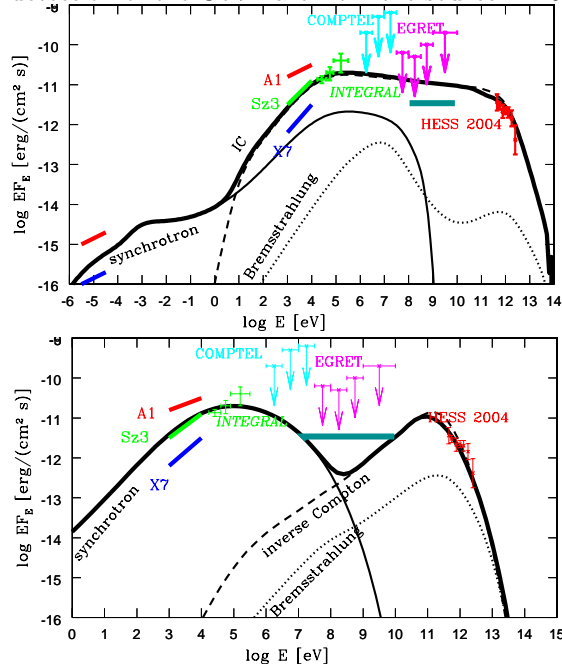}
\caption{Spectral energy distribution of \psrb\  in the IC  (top) and synchrotron (bottom) models (from Ref. \cite{chernyakova09}).  Thick cyan line shows estimated 1 month Fermi sensitivity.}
\label{models}
\end{figure}

Similarly to the case of LSI +61 303, physical mechanism through which X-ray to \gr\ emission from the source is produced (synchrotron or inverse Compton), is not well constrained \cite{chernyakova06,tavani97, khangulyan07,uchiyama09}.  In fact, spectral and timing characteristics of the source could be successfully reproduced in both models, see Fig. \ref{models}. 

From Fig. \ref{models} it is clear that {\it Fermi} observations of the next flaring period of the source during 2010 periastron passage will  clarify the mechanism of X-ray-to-\gr\ emission. Predictions of the two models significantly differ in the two cases. If the X-ray-\gr\ emission is dominated by inverse Compton emission from single population of electrons, the source should be detected by {\it Fermi} during the several month-long flare around the  periastron. The source spectrum is expected to be a powerlaw. At the same time, the source could remain below {\it Fermi} sensitivity level if the the X-ray emission is dominated by synchrotron. The cut-off energy of the synchrotron spectrum is not currently constrained by the available data. If the energies of electrons accelerated at the interface of pulsar and stellar wind (or injected directly from the pulsar wind) reach $\sim 100$~TeV energies (similarly to the case of \lsi), the high-energy cut-off of the synchrotron spectrum might be revealed by {\it Fermi}. 

\section{Summary}

We discussed the results of {\it Fermi} observations of GRLBs. We found that {\it Fermi} data significantly constrain the mechanisms of \gr\ emission from \lsi\ and enable to work out firm predictions for the expected neutrino signal from the source in the framework of hadronic models of source activity.  For \psrb\ we discussed the possibility of detection of the GeV flare from the source in 2010, during the period of passage of the young pulsar in the system close to the massive star. 




\bigskip 

\end{document}